# Resonant versus non-resonant spin readout of a nitrogen-vacancy center in diamond under cryogenic conditions

Richard Monge[1,2,*], Tom Delord[1,*], Gergő Thiering[3], Ádám Gali[3,4], and Carlos A. Meriles[1,2,†]

*[1]Department. of Physics, CUNY-City College of New York, New York, NY 10031, USA.*
*[2]CUNY-Graduate Center, New York, NY 10016, USA.*
*[3]HUN-REN Wigner Research Centre for Physics, P.O. Box 49, H-1525 Budapest, Hungary.*
*[4]Department of Atomic Physics, Institute of Physics, Budapest University of Technology and Economics, Műegyetem rakpart 3., H-1111 Budapest, Hungary.*

*[*]These authors contributed equally to this work.*

*[†]Corresponding author. E-mail: cmeriles@ccny.cuny.edu.*

The last decade has seen an explosive growth in the use of color centers for metrology applications, the paradigm example arguably being the nitrogen-vacancy (NV) center in diamond. Here, we focus on the regime of cryogenic temperatures and examine the impact of spin-selective, narrow-band laser excitation on NV readout. Specifically, we demonstrate a more than four-fold improvement in sensitivity compared to that possible with non-resonant (green) illumination, largely due to a boost in readout contrast and integrated photon count. We also leverage nuclear spin relaxation under resonant excitation to polarize the $^{14}$N host, which we then prove beneficial for spin magnetometry. These results open opportunities in the application of NV sensing to the investigation of condensed matter systems, particularly those exhibiting superconducting, magnetic, or topological phases selectively present at low temperatures.

The ability to initialize and read out the spin state of individual magneto-optically active color centers is presently driving extensive work aimed at their use as local probes to sense temperature, pressure, or electric and magnetic fields with spatial resolution down to the nanoscale[1]. Among a rapidly growing palette of material systems — including, e.g., color centers in silicon carbide[2,3] and hexagonal boron nitride[4,5] — the nitrogen–vacancy (NV) center in diamond stands out as a leading sensing platform due to a favorable combination of fluorescence brightness, readout contrast, and spin coherence lifetime.

While ambient operation has been key to their success, bringing NVs to cryogenic conditions creates opportunities for control schemes not possible at room temperature. Indeed, orbital averaging of the NV first excited manifold — fast at 300 K — effectively vanishes below ~20 K[6,7] leading to a set of narrow, spin selective optical transitions, already exploited, e.g., in demonstrations of spin-photon[8] and photon-mediated, inter-spin entanglement[9]. NV sensing protocols adapted to the cryogenic regime are less developed but equally attractive, particularly given the growing number of studies centered on condensed matter phases selectively present at low temperatures[10-14]. For example, recent work used resonantly tuned laser light between the ground and first excited NV manifolds at ~4 K to attain single-shot readout fidelity through efficient spin-to-charge conversion without resorting to high magnetic fields or photon-collection optics[15,16].

Despite their superior fidelity, spin-to-charge methods are not necessarily optimal because their intrinsically longer readout times — typically a few milliseconds — carry a concomitant overhead, countered only in sensing protocols demanding extended evolution intervals (such as, e.g., spin-lattice relaxation measurements[17-19]). On the other hand, standard (i.e., non-resonant) optical readout schemes tend to work more poorly at lower temperatures because electric-field- and strain-induced excited state anti-crossings — affecting individual NVs differently — are often detrimental to spin initialization and readout[20].

Here, we investigate NV spin readout under resonant laser light in the regime of low temperatures and low magnetic fields. We first characterize the NV spin dynamics in the presence of tunable, narrow-band optical excitation, and make use of spin initialization and readout schemes with minimal overhead to attain a multi-fold detection sensitivity boost (typically ~4 times, depending on the chosen NV) compared to non-resonant methods. We then turn our attention to the interplay between the NV and the $^{14}$N host, and show that resonant illumination alters the electronic and nuclear spin states on comparable time scales. We capitalize on this finding to polarize the $^{14}$N spin without resorting to radio-frequency (RF) pulses, which we subsequently show to be advantageous for applications to magnetic sensing.

*Optical spectroscopy at cryogenic temperatures.* Throughout our experiments, we investigate naturally occurring NV centers in a [100] electronic grade diamond using a homemade, multicolor confocal microscope based on a close-cycle cryo-workstation[21,22]. The system features green and red laser light sources (respectively, 532 and 637 nm); the latter is narrow-band (500 kHz) and tunable over a 4 nm range. We use half- and quarter-wave plates to control both the red beam plane of polarization and helicity, and collect the Stokes-shifted NV fluorescence using a 650-nm, high-pass filter and a single-photon avalanche photodetector (APD). We attain ground state control of the NV spin through the microwave (MW) field produced by an antenna in the form of a thin copper wire overlaid on the diamond surface.

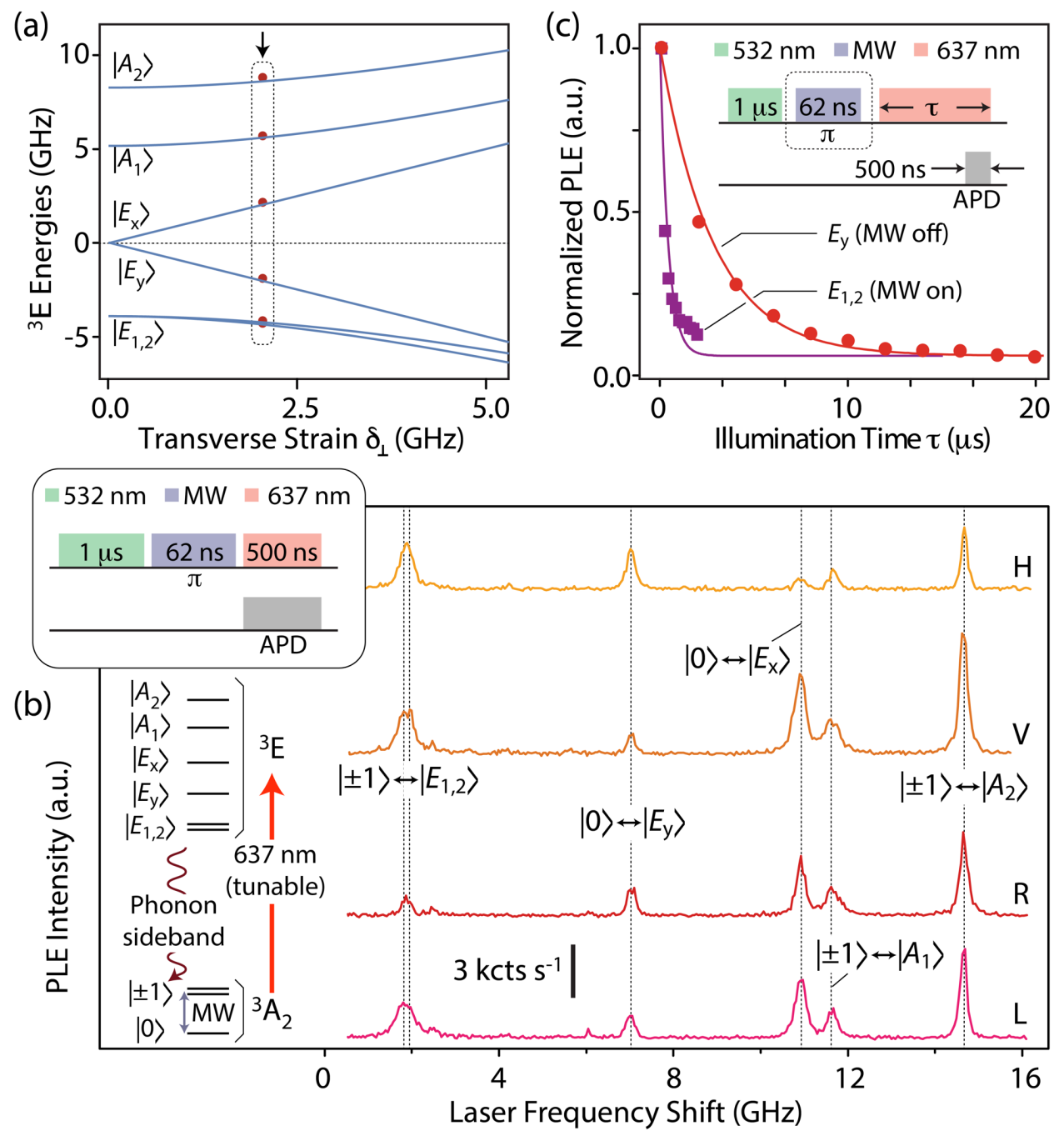

**FIG. 1: Resonant spectroscopy of NV centers.** (a) $^3$E energy diagram as a function of transverse strain. (b) PLE spectroscopy of a representative single NV; peak positions are signaled in (a) by the boxed dots. V and H indicate linear polarization of the probe beam along perpendicular planes; L and R denote opposite circular polarization helicity. The MW $\pi$-pulse makes $m_S = \pm1$ transitions visible (even if the amplitude of the $m_S = 0$ resonances is less than optimal). Frequencies are relative to 470.488 THz; spectra have been vertically offset for clarity. (c) PLE amplitude as a function of the red illumination duration under a L-polarized beam for the $E_y$ and $E_{1,2}$ transitions; solid lines are exponential fits with time constants $3.11 \pm 0.08$ μs and $0.19 \pm 0.02$ μs, respectively. The green (red) laser power is 500 μW (4 μW), and the MW frequency — resonant with the ground state $|0\rangle \leftrightarrow |-1\rangle$ transition — is 2.809 GHz; the temperature is 5 K and we apply no external magnetic field.

Figure 1a lays out the calculated NV $^3$E excited state manifold[22-33]: As an orbital doublet and spin triplet, it features six different states whose relative alignment and degeneracy depend on local strain and electric field. In our experiments, we tune the red laser to induce transitions between the $^3$E and the ground state, $^3A_2$, also a spin triplet but an orbital singlet whose response to local fields is weaker and has been ignored. Figure 1b shows example photo-luminescence excitation (PLE) spectra for a low-strain ($\delta_\perp \approx 3$ GHz) bulk NV under variable laser light polarization[22]. Except for $|E_x\rangle$ and $|E_y\rangle$ — primarily associated to the $m_S = 0$ spin projection — most eigenstates in the $^3$E manifold are entangled states of spin and orbital angular momentum[34]. Correspondingly, resonant optical excitation produces a transition-dependent decline of the observed fluorescence due to gradual spin depletion[35]; the decay rate is substantially smaller for the $|E_x\rangle$ and $|E_y\rangle$ resonances, which we hence refer to hereafter as the 'cycling' transitions (Fig. 1c).

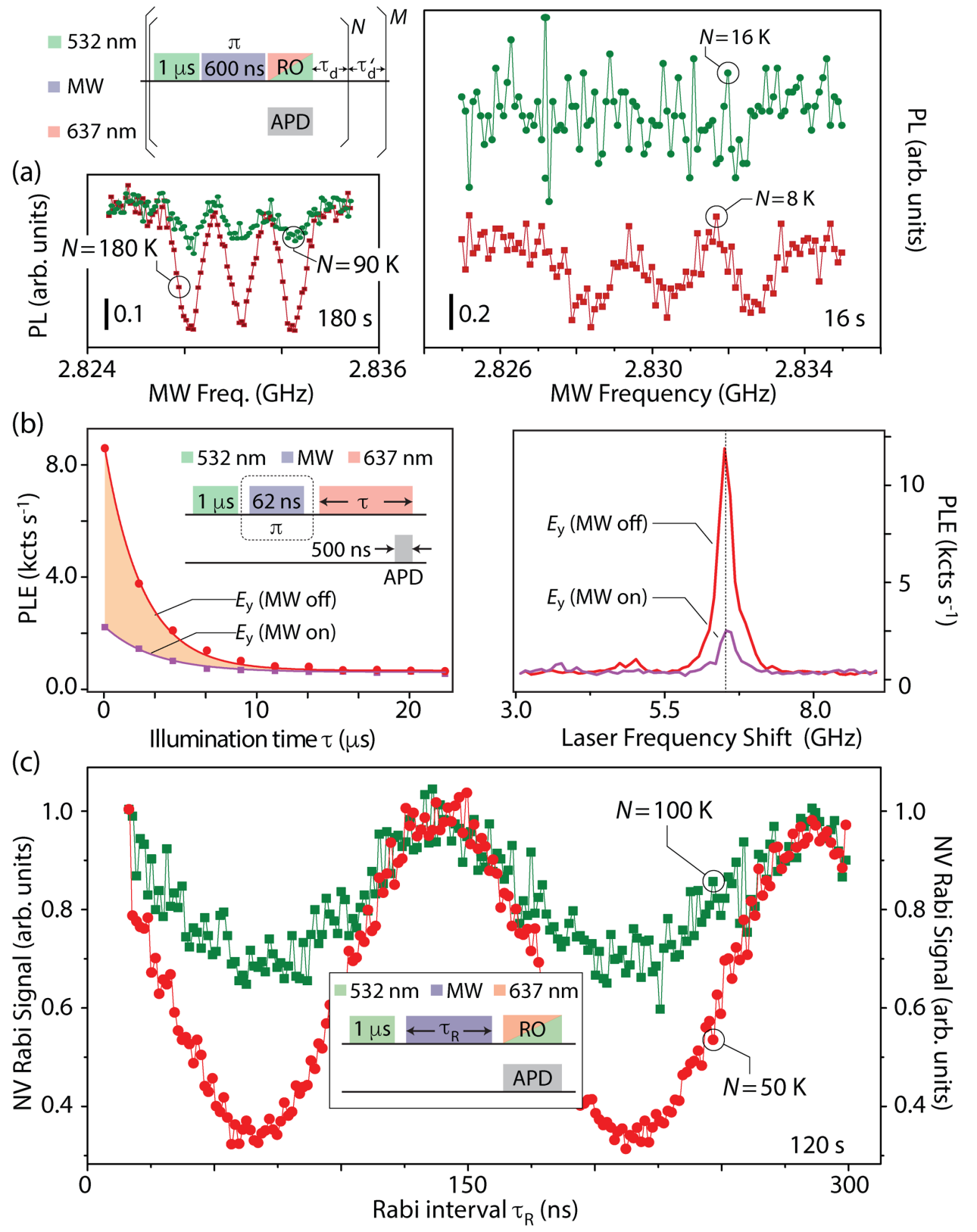

**FIG. 2: Comparison of NV resonant and non-resonant readout.** (a) (Top left) Pulsed ODMR using red (637 nm) or green (532 nm) readout (RO); $\tau_d$ and $\tau_d'$ are hardware-imposed delays. (Right) ODMR for the NV in Fig. 1 using green or red readout (upper and lower traces, respectively); in both cases, the number of sampled frequencies is $M = 101$ and the experimental time is 16 s (Bottom left) Same as before but for an experimental time of 180 s. (b) (Left) PLE amplitude for the $E_y$ transition with or without MW inversion; resonant readout results from the integrated photon count difference (shaded area). (Right) $E_y$ photoluminescence as a function of laser frequency using the protocol in the left plot but with a fixed red illumination time (0.5 μs). (c) NV spin Rabi signal for resonant and non-resonant readout (red circles and green squares, respectively) as a function of the MW pulse duration for a common 120 s experiment time; in (a) and (c), the red (green) readout time is 5 μs (0.3 μs). In all these experiments, the red (green) laser power is 10 μW (500 μW) and the magnetic field — ~5 deg misaligned with the NV axis — is ~2 mT.

*NV readout under optically selective excitation.* A key metric for applications of single spin-active color centers to sensing is given by the sensitivity $\eta$, expressed herein as

$$\eta = \frac{\mathcal{A}}{\sqrt{N t_{\mathrm{E}}}\, e^{-\frac{t_{\mathrm{E}}}{t_{\mathrm{C}}}}}\sqrt{\left(1+\frac{1}{\mathcal{C}^2 n_{\mathrm{avg}}}\right)\left(\frac{t_{\mathrm{I}}+t_{\mathrm{R}}+t_{\mathrm{E}}}{t_{\mathrm{E}}}\right)}, \quad (1)$$

where $\mathcal{C}$ is the effective spin contrast, $n_{\mathrm{avg}}$ is the net number of photons during the readout time $t_{\mathrm{R}}$, $t_{\mathrm{E}}$ denotes the 'evolution' (i.e., spin-signal-encoding) time, $t_{\mathrm{I}}$ is the system initialization time, and $t_{\mathrm{C}}$ is a sequence-specific NV spin coherence lifetime; $N$ represents the total number of repeats, and $\mathcal{A}$ is a constant that depends on the specific transduction mechanism connecting the color center spin to the external stimulus being measured (e.g., electric, magnetic, etc.).

Resonant NV spin readout is particularly favorable at a cycling transition because both the contrast and integrated photon counts are greater compared to those possible under standard green illumination. As a first illustration, Fig. 2a shows pulsed optically detected magnetic resonance (ODMR) data at 7 K for the NV in Fig. 1 using green and red ($E_y$) readouts; the protocol uses a green laser pulse to initialize the NV charge[36] (preferentially into negative) and spin[37] states (predominantly into $m_S = 0$). A clear readout sensitivity gain — approximately 4.3 times in this case — emerges under resonant excitation, largely a consequence of the greater integrated photon count ($n_{\mathrm{avg}}$ is $8.8\times10^{-2}$ per readout for red light but only $5.9\times10^{-3}$ for green[22]). Further, the readout contrast $\mathcal{C}$ — weaker in a pulsed ODMR experiment due to MW spectral selectivity — also experiences a boost under resonant illumination reaching up to 78%, limited by NV spin initialization[15,16] (Fig. 2b). The result is an enhanced detection sensitivity even in a routine sensing protocol like this one, with comparable — if not better — improvements expected in all other (more time-consuming) NV control schemes. Figure 2c displays an example in the form of NV Rabi responses using resonant and non-resonant readout; the signal-to-noise ratio (SNR) boost is substantial, here exposed upon imposing a common experimental time of 120 s. We have observed comparable — if slightly smaller — sensitivity gains for other NVs throughout the crystal[22].

*Polarization of the nuclear spin host.* An open question relevant to NV cryo-sensing applications is the impact of selective optical excitation on the nuclear spin polarization of the nitrogen host. Specifically, we are interested in determining the time scale over which resonant illumination alters the starting nuclear spin state. We address this problem in Fig. 3, where we first initialize the NV spin into $|m_S = -1\rangle$ using a population trapping protocol comprising sequential optical excitation and

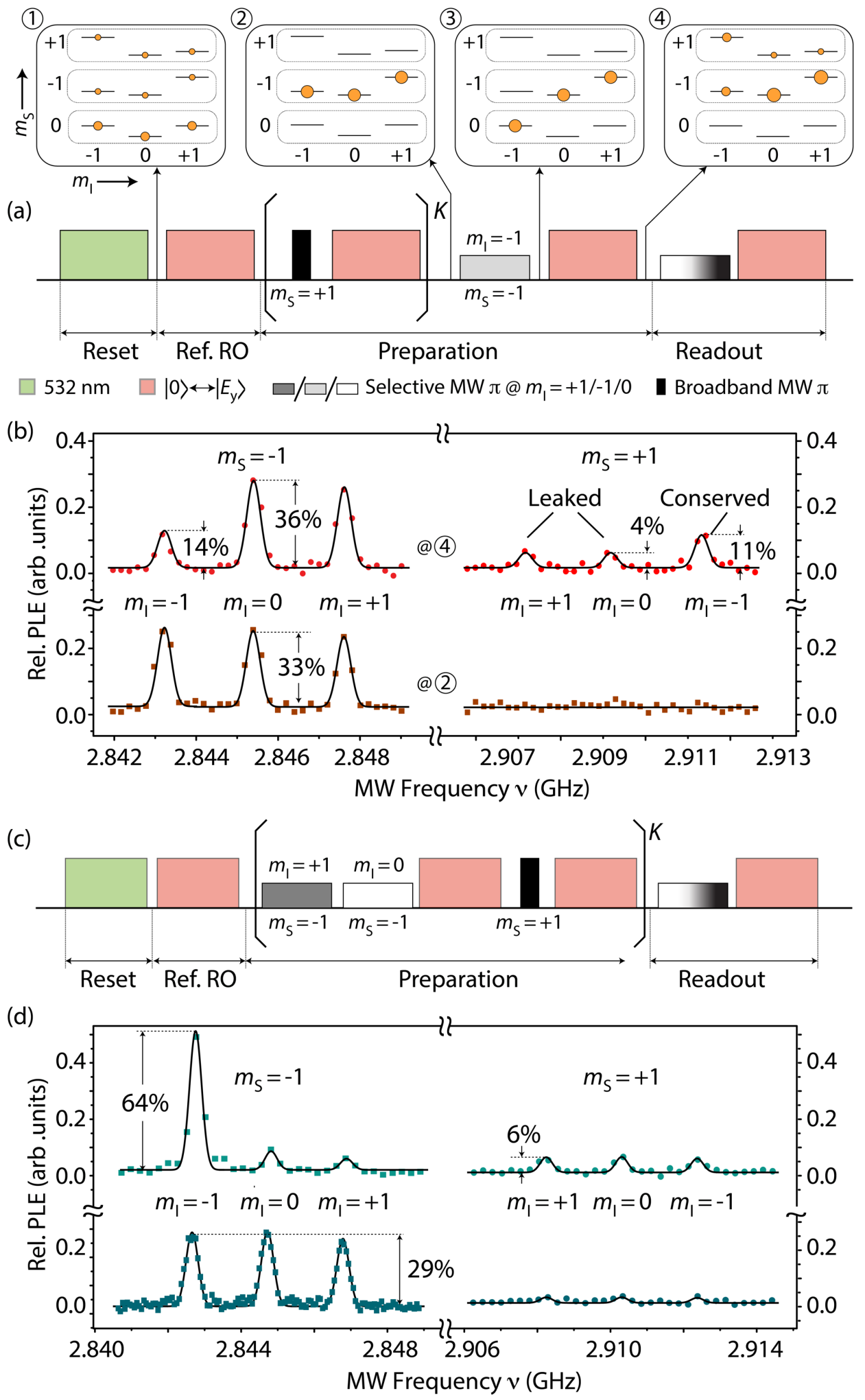

**FIG. 3: RF-free spin polarization of the $^{14}$N host.** (a) To study the action of resonant excitation on the $^{14}$N nuclear spin state, we first initialize the electronic spin state into $m_S = -1$ manifold. Upon selective MW inversion of the $|m_S = -1, m_I = +1\rangle$ state and resonant excitation of the $|E_y\rangle$ transition, we probe the hyperfine populations in the $m_S = +1$ manifold. (b) NV ODMR spectrum after application of the protocol in (a) using $K = 5$; the lower data set shows the background signal obtained in the absence of a selective $\pi$-pulse in the $m_S = -1$ manifold. (c) $^{14}$N nuclear spin polarization protocol. (d) NV ODMR spectrum upon application of the protocol in (c) for $K = 6$; the lower traces are the spectra in the absence of selective $m_S = -1$ pulses. In (a) and (c), we use a color gradient to denote a selective MW $\pi$ -pulse whose frequency we vary for spectrum reconstruction. In (b) and (d), percent values indicate the estimated fractional nuclear spin populations with errors within ±2%; solid lines are Gaussian fits. The red (green) laser power is 1 μW (100 μW) and the MW pulses are 2.4 μs long. Ref. RO: Reference readout.

electronic spin state inversion[15]. Upon selectively populating one of the hyperfine levels in the $|m_S = 0\rangle$ manifold (specifically, the $|m_S = 0, m_I = -1\rangle$ state in Fig. 3), we determine the optically induced change in the nuclear spin state by reconstructing the NV ODMR spectrum following resonant optical spin depletion of

$|m_S = 0\rangle$ (see schematics in Fig. 3a).

Figure 3b shows the results: Comparison of the relative hyperfine peak amplitudes — correlating with the nuclear spin populations after resonant optical excitation — shows that a single cycle of electronic spin depletion has a significant probability (approximately 30%) to simultaneously flip the nuclear spin state, even in the case of a cycling transition such as the $|m_S = 0\rangle \leftrightarrow |E_y\rangle$ probed here. While hyperfine states are known to be fragile against green optical excitation at low magnetic fields[38], the above result is surprising in that the $|E_y\rangle$ state has a predominant $m_S = 0$ character[34], and hence one would anticipate a reduced hyperfine coupling or, equivalently, a slow nuclear spin flip rate. Further, we find that starting from $|m_I = -1\rangle$, both $|m_I = 0\rangle$ and $|m_I = +1\rangle$ populate comparably under resonant illumination, thus pointing to efficient double-quantum nuclear spin relaxation processes. A detailed analysis suggests single- and double-quantum flips stem from complementary channels, associated to transverse hyperfine and quadrupolar couplings, respectively[22].

While a fuller understanding of the dynamics at play will require additional work, one important implication for NV sensing is that one can combine resonant illumination and selective MW excitation to polarize the nuclear spin state of the nitrogen host without resorting to radio-frequency pulses[39]. We illustrate this notion in Figs. 3c and 3d where we modify the protocol in Fig. 3a to trap the NV population into the $|m_S = -1, m_I = -1\rangle$ state prior to optical spin readout. From an analysis of the resulting ODMR spectrum, we conclude the spin trapping efficiency here reaches up to 64%, limited by imperfect spin depletion of the $m_S = 0$ manifold[22].

Interestingly, the above nuclear spin polarization scheme is not unique as other, less time-consuming routes are also possible, though at the expense of a reduced polarization efficiency. To illustrate this point, we return to the regular ODMR protocol in Fig. 2a, though this time we extend the duration of the MW pulses (2.4 μs) to attain good spectral resolution (red trace in the upper plot of Fig. 4a). Remarkably, we find that the contrast we attain via selective MW excitation (reaching up to 39%) exceeds that possible for an NV center whose nuclear spin host is unpolarized (~23%), as we confirm by comparing to the ODMR signal under 80-ns-long π-pulses (featuring ~70% contrast, see brown trace in the lower plot of Fig. 4a). An in-depth analysis[22] shows this response stems from a measurement-induced process during the ODMR sequence, where multiple repeats of resonant optical and MW excitation combine to partially polarize the nuclear spin host. Note that the contrast we attain under selective MW excitation (smaller than in Fig. 3d but greater than the best possible under green readout, see faint green trace in the lower plot of Fig. 4a) depends delicately on the duration of the 532-nm initialization pulse, consistent with the scrambling action of green illumination on the nuclear spin[38] (see faint pink trace in the upper plot of Fig. 4a).

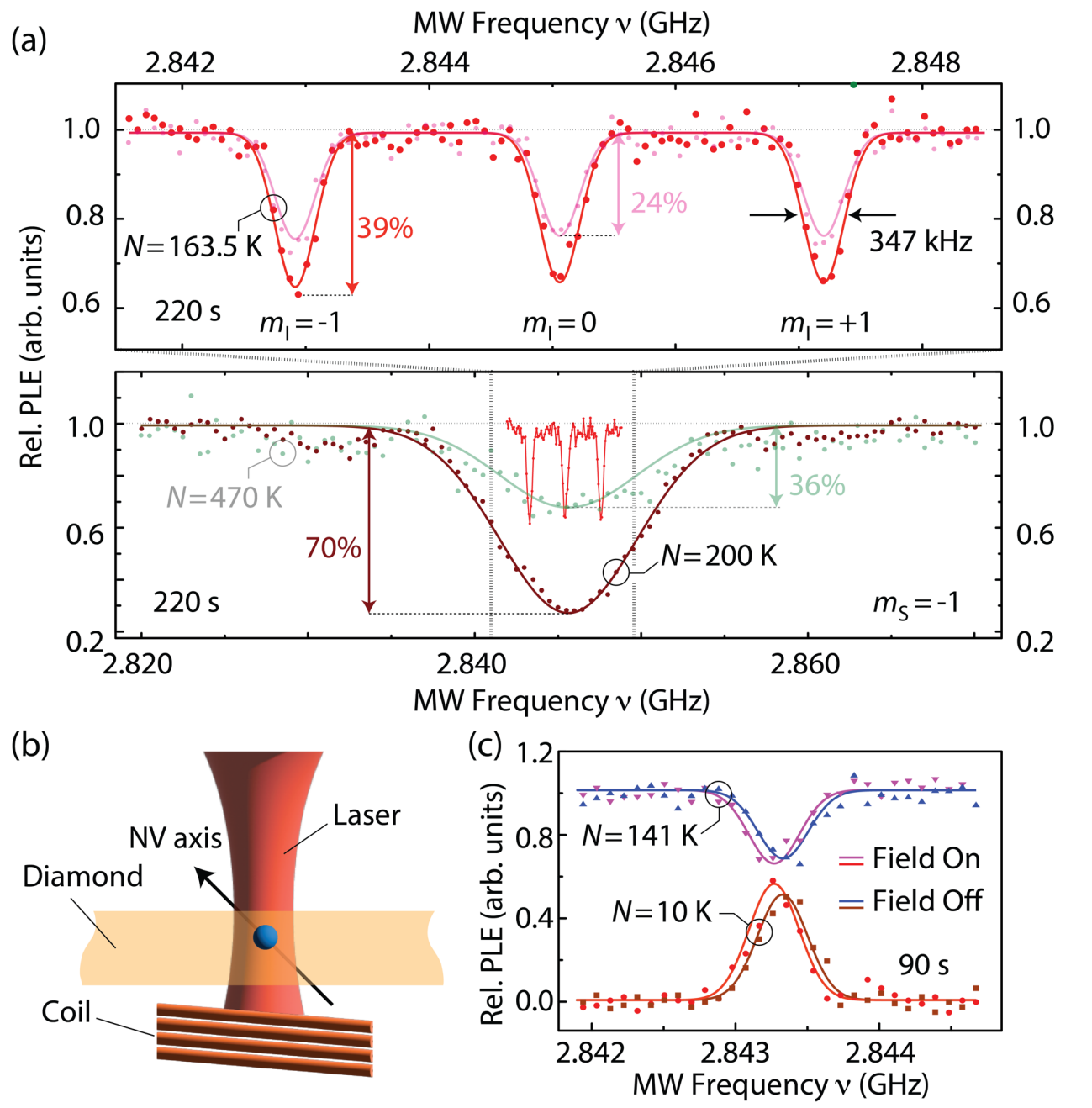

**FIG. 4: Magnetometry with nuclear-spin-polarized NVs.** (a) (Bottom plot) NV ODMR spectroscopy using the protocol of Fig. 2a using a 5-μs red readout and selective (2.4 μs) or broadband (89 ns) π-pulses (red and brown traces, respectively). For reference, the faint green trace in the back shows the NV ODMR spectrum under standard green readout (300 ns). In all cases, the number of sampled frequencies is $M = 101$, the experimental duration is 220 s, and the green initialization pulse is 2 μs. (Upper plot) Zoomed NV ODMR spectrum under selective MW excitation. The faint pink trace shows the same result but for 5-μs-long green initialization pulses, resulting in reduced contrast. (b) Schematics of the experimental setup. (c) NV ODMR spectra with or without a coil-generated, ~3 μT magnetic field. The upper traces use the conditions in Fig. 4a (red set) to induce partial nuclear spin polarization; in the lower traces, we apply the protocol in Fig. 3c. In (a) and (c), solid lines are Gaussian fits.

*Magnetometry with a spin-polarized host.* The ability to initialize the nuclear spin host translates into a higher sensitivity to magnetic field shifts. We experimentally demonstrate this notion in Fig. 4b where we use a small coil adjacent to the diamond crystal to controllably change the external magnetic field by a small amount. Interestingly, the "direct current" (DC) magnetic sensitivity we attain via a measurement-induced $^{14}N$ polarization ($\eta_{DC}^{(MI)} = 1.74 \pm 0.50$ μT Hz$^{-1/2}$, upper traces in Fig. 4c) is comparable to that resulting from resorting to the active nuclear polarization protocol of Fig. 3c ($\eta_{DC}^{(AP)} = 0.90 \pm 0.12$ μT Hz$^{-1/2}$, lower traces in Fig. 4c), largely due to a much reduced overhead. Comparing to the sensitivity possible in our set-up via green read-out, the results above correspond to 6- or 7-fold enhancements for this NV; we find comparable — if slightly smaller — improvements for others, and expect as high as ten-fold boosts for longer NV coherence[22].

In summary, we have shown how resonant optical excitation under cryogenic conditions can improve NV spin detection sensitivity over standard green readout. At the low magnetic fields present in our experiments, 637-nm light induces unexpectedly fast nuclear spin flips of the nitrogen host, a finding we exploited to demonstrate efficient $^{14}$N polarization. Combining resonant excitation and nuclear spin initialization, we demonstrated improved DC magnetic field sensitivity. Note that $^{14}$N spin polarization should also have a positive impact on the attainable sensitivity to "alternating current" (AC) magnetic fields, not only due to the greater contrast but also because the narrower spectral bandwidth makes spin decoupling protocols less sensitive to frequency offset. Future work must address the transition from bulk to shallow NVs as surface proximity is known to be detrimental to the cyclicity and brightness of the optical transitions (although the reasons are only partly understood). This effort should prove valuable in the growing set of applications of NV sensing to solid state phases selectively present at low temperatures.

All authors acknowledge support from the National Science Foundation through grants NSF-2216838, NSF-2203904, and NSF-1903839. R.M. acknowledges support from NSF-2316693. A.G. acknowledges the support from the NKFIH in Hungary for the National Excellence Program (Grant No. KKP129866), the Quantum Information National Laboratory (Grant No. 2022-2.1.1-NL-2022- 0000), and the EU QuantERA II MAESTRO project; he also acknowledges support from the European Commission through the QuMicro project (Grant No. 101046911). A.G. and G.T. acknowledge the high-performance computational resources provided by KIFÜ (Governmental Agency for IT Development) Institute of Hungary. All authors also acknowledge access to the facilities and research infrastructure of the NSF CREST IDEALS, grant number NSF-2112550.

# Supplementary Material for

# "Resonant versus non-resonant spin readout of a nitrogen-vacancy center in diamond under cryogenic conditions"

Richard Monge[1,2,*], Tom Delord[1,*], Gergő Thiering[3], Ádám Gali[3,4], and Carlos A. Meriles[1,2,†]

[1]*Department. of Physics, CUNY-City College of New York, New York, NY 10031, USA.*
[2]*CUNY-Graduate Center, New York, NY 10016, USA.*
[3]*HUN-REN Wigner Research Centre for Physics, P.O. Box 49, H-1525 Budapest, Hungary.*
[4]*Department of Atomic Physics, Institute of Physics, Budapest University of Technology and Economics, Műegyetem rakpart 3., H-1111 Budapest, Hungary.*

[*]*Equally contributing authors*
[†]*Corresponding author. E-mail: cmeriles@ccny.cuny.edu..*

## 1-Experimental

We work with two electronic-grade [100] single-crystal diamonds. The samples are mounted on a copper holder using a small amount of silver paint so as to ensure good thermal contact without introducing crystal strain. Throughout our experiments, we use a home-built confocal microscope featuring a green laser at 532 nm and a tunable, 150-kHz-linewidth red laser at 637 nm (Toptica DL pro HP 637), which we stabilize (±200 MHz absolute accuracy) with a wavemeter (High-Finesse WS/6-200). We merge the laser beams into a single mode fiber, and produce laser pulses down to 5 ns duration with the aid of two independent acousto-optic modulators (AOMs); we detect the NV fluorescence via a single-photon avalanche photodetector (APD) from Excelitas. Long-pass filters and a 650-nm dichroic mirror allow us to reject laser excitation leaking into the detection path. We use a cryo-workstation (Montana Instruments) to reach low-temperatures down to 4 K. High power application of microwave excitation warms the samples $< 2$ K. Inside the vacuum chamber sits an air objective (100× Zeiss Epiplan Neofluar) with a numerical aperture of 0.75.

A permanent magnet creates a magnetic field of 1.2 mT nearly aligned with the NV axis (less than 5 deg.), as derived from a comparison of the $|m_S = 0\rangle \leftrightarrow |m_S = \pm 1\rangle$ transition frequencies[1]. To control the NV electronic spin, we use microwave (MW) excitation from four signal generators (Rhode&Schwarz SMB100A ×3 and a Stanford Research Systems SG386). We create MW pulses with the help of fast switches from Minicircuits; all MW signals are combined using a 3-port splitter/combiner (ZB2PD-63-S+) and a 2-port splitter/combiner (ZAPDQ-4-S+). After amplification (Minicircuits ZHL-16W-43-S+), we drive the MW into a 25-μm-diameter uncoated copper wire overlaid on the sample, which we use as a local antenna.

| NV | MW excitation | HP | π-pulse Duration (μs) | RO laser | Sensitivity (μT/Hz$^{1/2}$) | Sensitivity Enhancement |
|---|---|---|---|---|---|---|
| A | Selective | MI | 2.2 | Red | 0.90±0.12 | 6.14±2.2 |
| A | Selective | AP | 2.2 | Red | 1.02±0.26 | 7.0±2.5 |
| A | Selective | None | 2.2 | Red | 1.74±0.50 | 3.6±1.2 |
| A | Selective | None | 2.2 | Green | 6.27±2.22 | 1 |
| A | Broadband | None | 0.06 | Red | 12.8±1.3 | 5.3±0.6 |
| A | Broadband | None | 0.06 | Green | 68.1±7.5 | 1 |
| B | Selective | MI | 1.2 | Red | 1.29±0.11 | 4.3±0.6 |
| B | Selective | None | 1.2 | Green | 5.49±0.70 | 1 |
| C | Selective | MI | 2 | Red | 3.34±0.28 | 3.6±0.4 |
| C | Selective | None | 2 | Green | 12.05±1.35 | 1 |
| D | Selective | MI | 1.2 | Red | 1.705±0.15 | 4.0±0.5 |
| D | Selective | None | 1.2 | Green | 6.851±0.83 | 1 |

**Table 1:** Comparison of magnetic sensitivities for red and green readout, under broadband or selective (i.e., narrow band) MW excitation and for several NVs with various strain splitting $\delta_\perp$ (from A to D: 7, 0.5, 12 and 21 GHz). The red (green) readout duration is 2-7 μs (300 ns). HP: hyperpolarization, MI: Measurement Induced, AP: Active Polarization.

The experiments are controlled via a PulseBlaster ESR PRO card and the CR acquired by an NI PCIe6321 card. Delays can occur between two experiments due to software processing time (~25 ms), sequence loading in the Pulse Blaster (~50 ms), hardware communication (~25 ms), and MW generator settling time (~5 ms). In addition, small delays occur for each iteration of the acquisition sequence to allow for AOM fall/rise time (15-50 ns) and delayed action (500ns-2μs), MW ring-down (20-100 ns) and singlet state relaxation (1 μs). The total delays as defined in Fig. 2 of the main text are $\tau'_d \sim$ 50-100 ms and $\tau_d \sim$1.5-3 μs depending on the level of optimization and sequence type.

Data acquisition starts with charge and spin initialization using a 600 ns to 10 μs green laser pulse (depending on power, typically 600 ns at 1.5 mW) followed by a 500 ns-1 μs wait time to allow for relaxation from the singlet metastable state. PLE spectra are carried out with short red illumination (typically 500 ns) following green initialization and preceded by an additional MW π-pulse for $m_S = \pm 1$ initialization.

## 2-Measurement of the magnetic sensitivity

We determine the magnetic sensitivity for green and red RO using the pulsed ODMR method [1]. For each NV, we typically choose the optimal value for the π-pulse duration, equal to the $T_2^*$ [1]. We determine $T_2^*$ by performing a Ramsey sequence and fitting the response to damped sinusoidals characterized by the detuning of the excitation MW frequency. Below we consider four example NVs, referred to as NV A, B, C or D. We use 2.4 μs for NV A, 1 μs for NV B, and 2 μs for NV C. For NV B, C and D and for each read-out wavelength, we also optimize the laser powers and RO windows following a protocol described in the following sections. RO windows are typically 300 ns for green and 2 to 7 μs for red.

We calculate the magnetic sensitivity for a field along the NV axis using pulsed-ODMR swept in frequency. The specific sequence we use is typically repeated a few thousand times for every frequency of the sweep, with a typical acquisition time $T$ per point in the range of seconds. To estimate the signal amplitude in a given sequence, we measure two ODMRs for slightly different fields by applying a 0-1 mA current in a loop tucked under the diamond sample. We then fit those ODMR signals to determine first the magnetic field shift $\delta B$ generated by the loop, and second the PL change $\delta S$ by taking the difference between the two fitted curves at the point of maximum slope. The noise is then estimated by calculating the root mean square deviation ($RMSD$) of the experimental data from the fit. The sensitivity $\eta$ in $\mu T/\sqrt{Hz}$ is finally obtained by applying correction factors to the field we measure with the formula $\eta = \delta B.\sqrt{T}.RMSD/\delta S$. Note that this estimation relies on oversampling the frequency of the sweep to accurately measure the noise, and that too few points would introduce errors in the noise estimation. Alternatively, the sensitivity can also be derived from a single ODMR using a similar method but taking the maximum value of the slope to calculate the expected signal $\delta S$ for a given small field $\delta B$ [1,2].

Table 1 shows a summary of the measured sensitivity under green and red excitation, both under broadband or selective microwave excitation. Note that green RO of NVA is degraded due to high background PL while both absolute sensitivities of NV C are degraded by the presence of a nearby strongly coupled spin, lowering the contrast by a factor of 2. While the absolute sensitivities vary depending on the NV quality ($T_2^*$, strain, background PL), we consistently find an enhancement under red read-out sensitivity.

**3- Optimization of the spin read-out**

In this manuscript, we focus on enhancing the sensitivity to DC magnetic fields using resonant excitation of the NV center instead of green excitation: Our figure of merit is the relative sensitivity, i.e., the ratio between the sensitivity under resonant and green excitation. In that light, we do not explore other well-known avenues to enhance both red and green "absolute" sensitivities, e.g., we work with low photon extraction (NA of 0.75 without a solid immersion lens or nanostructure), natural abundance of $^{13}C$ (limiting the $T_2^*$ to hundreds of kHz) and a single power cw green illumination for the NV charge and spin initialization. Similarly, control of the experiment is not optimized for magnetic sensitivity: multiple hardware and software delays slow down fast acquisitions but can typically be neglected for longer averaging. Since those are heavily dependent on our specific software/hardware integration and could be optimized out, we calculate our sensitivity by timing only the execution time of the experimental sequence, excluding most delays due to data processing, communication with devices and software execution.

An accurate measure of the relative red/green sensitivities, however, requires systematic optimization of the PL read-out for both wavelengths. Laser powers well above saturation power are chosen to maximize the NV count-rate (CR) while limiting background contributions. Figure S1 (a,d) shows the PL as a function of laser power under cw green and pulsed red excitation, a fit gives saturation count rate (CR) of 21 kcps (35 kcps) and saturation powers of 0.63 mW (94nW) for green (red) excitation. Note that the reduced background PL under weak resonant excitation, though not critical for this NV presents an additional advantage for magnetometry applications comprising other optically active system near the NV. Here, we choose green (red) powers entering the objective of 1.5 mW (1.5 µW). We then optimize the PL read-out windows using the evolution of the spin-dependent PL under green and red illumination (Figure S1 (a,e)). These measurements allow us to estimate the measurement contrast, average count per measurement and maximum shot noise. We then estimate the expected SNR for one second as a function of the RO window for a specific sequence using the following formula, obtained by dividing the spin signal by the maximum shot noise:

$$SNR = (N_+ - N_-)\frac{1}{\sqrt{N_+}}\frac{1}{\sqrt{t_I + T_E + t_R}},$$

where $N_+$ and $N_-$ are the average photon per RO for the $m_S = 0$ and $m_S = \pm 1$ respectively, $t_I$ is the spin initialization time (typically 600 ns), $T_E$ is the evolution time of the spin (here a $\pi$-pulse or a spin echo sequence) and $t_R$ is the RO time.

Figure S1 (c,f) shows the result for green and red RO. Here, we choose 300 ns for the green RO and 6.5 µs for the red RO. This optimization is most important for the red RO where the spin depletion rate depends on the transverse strain and magnetic field. Importantly, we can confirm the expected SNR ratio between red and green RO by estimating the DC magnetic field sensitivity using pulsed ODMR (see section 2) on that same NV: we measure a magnetic sensitivity ratio of 4.5 for an expected SNR ratio of 4.8.

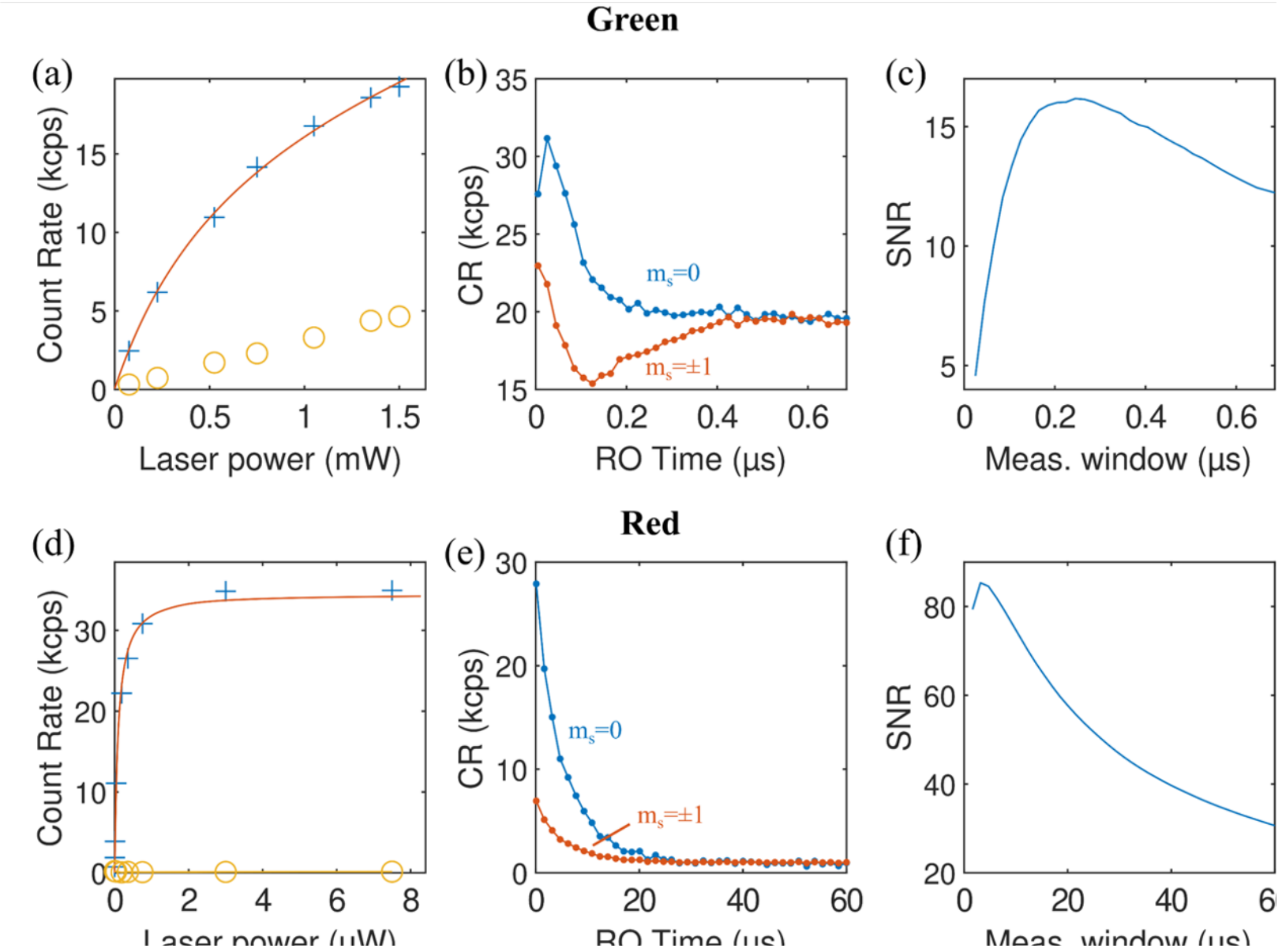

**Figure S1**: PL saturation under continuous green illumination (a) and pulsed resonant pulsed illumination (500 ns red pulse following a 1 µs green illumination) (b) traces (i) is the PL from the NV and background and (ii) from the background only. Evolution of the PL under 1.4 mW of green (b) and 1.5 µW of red (e) illumination after initializing in a specific spin state. Time constants for the spin signal exponential decay are 150ns and 5µs. Projected SNR for a 1s measurement as a function of the PL measurement window for green (c) and resonant (f) readout. We assume shot noise and take a sequence comprising RO as well as 1 µs relaxation time following a 600 ns green illumination and 1.2 µs of spin manipulation. These plots are all obtained from the $E_y$ transition of the low strain NV B (strain splitting ~0.5 GHz).

Note that for green RO, we typically include the RO time in the spin initialization time. For the red RO however, the optimization depends somewhat on $T_E$ with longer read-out time for longer sequences. For pulsed ODMR with $T_E = 1.2\ \mu s$, we take 300 ns (6 µs) windows for green (red) read-out.

## 4- Impact of the spin evolution overhead

The strong overhead incurred by the red RO becomes negligeable for longer measurement sequences, therefore increasing the sensitivity gain due to red RO. This can be seen by using the formula for DC magnetic sensitivity using pulsed ODMR:

$$\eta \cong \frac{8}{3\sqrt{3}} \frac{\hbar}{g_e \mu_B} \frac{1}{C\sqrt{N}} \frac{\sqrt{t_I + T_2^* + t_R}}{T_2^*},$$

where $N$ is the photon count per measurement, which increases with $t_R$. Since we take $T_E = T_2^*$ for optimal sensitivity [1], we see that the sensitivity gain under red RO ultimately depends on the coherence of the NV in question. $T_2^*$ can reach 500 µs in the bulk [3] and up to 149 µs (median at 26 µs) in a 110 nm thick diamond membrane [4], which is enough to overtake the red RO overhead. In AC magnetometry, the sequences duration scales as the inverse of the frequency, with longer duration leading to an increasing frequency resolution. To demonstrate the effect of red RO on sensitivity in those cases, we calculate the expected SNR as a function of the evolution time of our sequence $T_E$ using the method discussed in the previous section. Figure S2 (c) shows the red/green SNR ratio for increasing spin evolution time. The first red asterisk corresponds to the magnetic sensitivity gain in pulsed ODMR ($T_E = 1.2\ \mu s$). We see that for a coherence time of 26 µs we expect red RO with a similar NV to lead to a more than 10-fold improvement. We finally perform AC magnetometry using a spin echo sequence and compare the SNR under red and green RO (Figure S2 (a,b))for $T_E \cong 8\ \mu s$ obtaining an SNR ratio of 6.8, fairly close to the expected ratio of 8.4. We attribute the discrepancy to optimization of the RO window for DC magnetometry. Note that for long spin evolution times ($T_E \gtrsim 500$ µs), we eventually expect to enter a regime where spin to charge conversion [5,6] becomes more efficient than direct spin RO.

## 5- Impact of spectral diffusion

Spectral diffusion caused by change of the local charge environment can hinder efficient resonant read-out, which typically relies on the NV sharp optical transitions. The shallow NVs used for spatially resolved NV magnetometry are more susceptible to spectral diffusion due to the electric noise coming from the diamond surface and display inhomogeneous linewidth ranging from 160 MHz to several GHz [5,7–9]. Experiments with NVs implanted at select depth seem to indicate sharper lines for deeper NVs [5,7], while shallower NVs show a wider distribution of linewidth with a median linewidth as high as a few GHz [8]. Figure S3 shows the impact of spectral diffusion on the linewidth and saturation for NVs in the bulk: (a,b) shows the PLE spectrum for low strain NV B and another NV with higher strain and spectral diffusion, NV C. For each of the NVs, we measure the linewidth (Fig. S3 (c,d)) and maximum count rate (Fig. S3 (e,f) and S1 (d)) as a function of power by fitting PLE spectra at varying power with a Lorentzian. At high power, the linewidth increases due to power broadening while at low power it tends towards the NV linewidth, a convolution of the lifetime limited linewidth (~13 MHz) and the inhomogeneous broadening caused by spectral diffusion. NV B displays a low power linewidth of around 60 MHz while NV C stays above 400 MHz, demonstrating a high degree of spectral diffusion. This broadening can however be compensated by power broadening and the CR of NV C as a function of power still shows clear saturation with negligeable background for both $E_x$ and $E_y$.

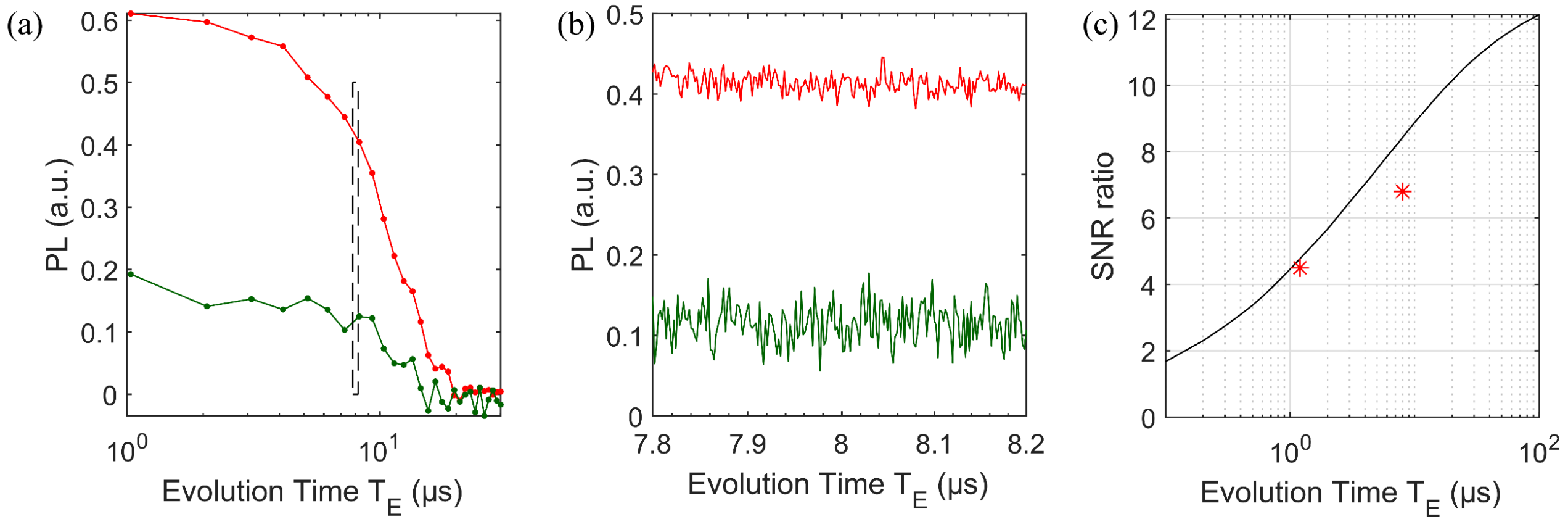

**Figure S2**: (a) Spin echo signal under red (top) and green (bottom). (b) Estimation of the spin echo noise around $T_E = 8$ µs. (c) Ratio of the expected spin RO SNR under red and green RO as a function of the spin evolution time. The two red asterisks are experimental values for the ratios of the magnetic sensitivity or SNR for DC magnetometry (left) and spin echo (right). Measurements are performed on NV B with green (red) laser powers of 1.5 mW (2 µW) and otherwise similar parameters as in Fig. S1.

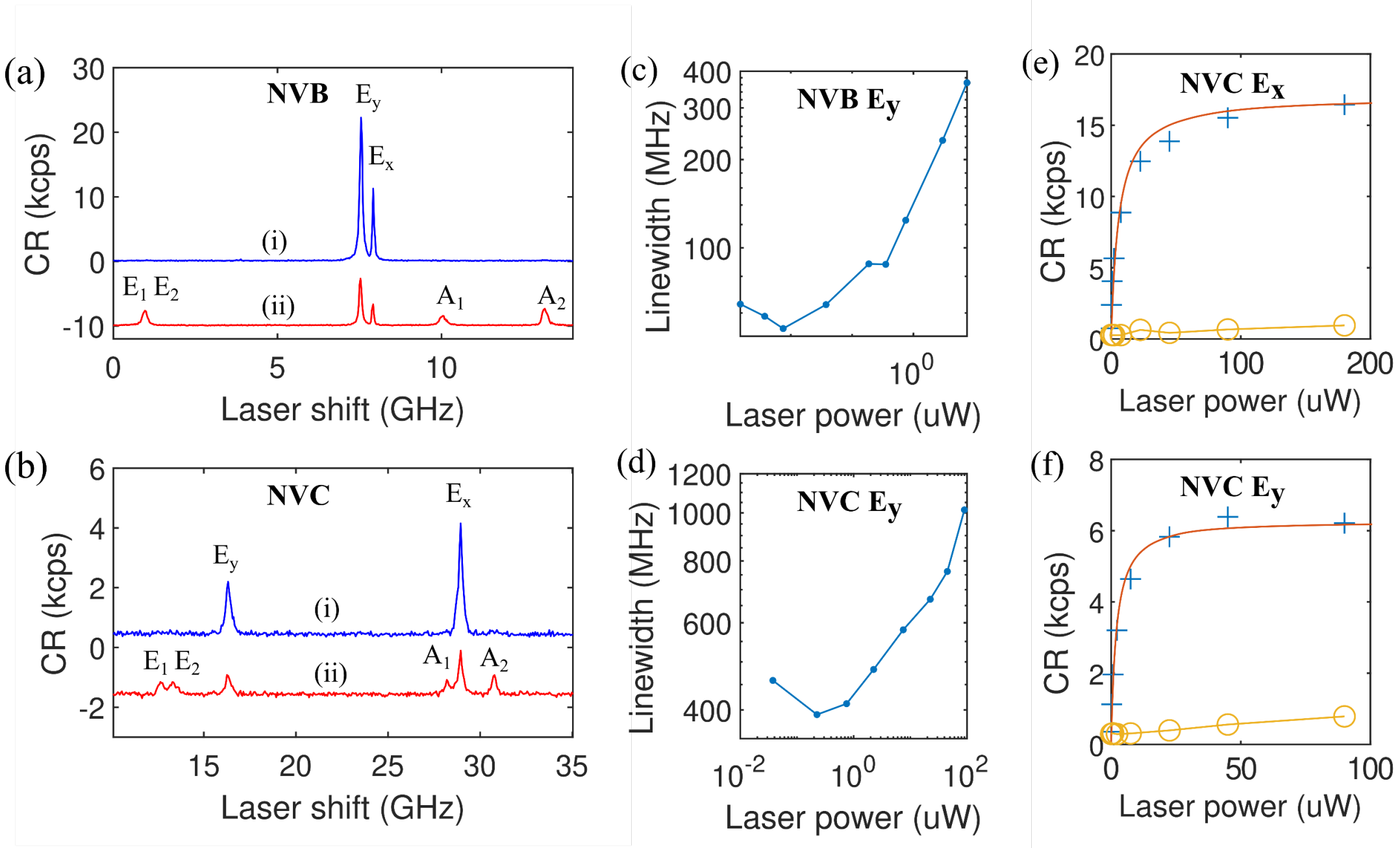

**Figure S3**: (a,b) PLE spectra of NV B (a) and NV C (b) (i) after green pump (ii) and a microwave π-pulse (displaced for visibility). RO is performed under 0.75 μW for 500 ns. Laser reference is 470.470 THz. (c,d) Linewidth as a function of power obtained from the PLE spectrum, using a Lorentzian fit. (e,f) CR of the peak amplitude (crosses) and background (circles) as a function of the laser power for the $E_x$ and $E_y$ transitions of NV C. Saturation count rates and powers are 17 kcps (6.4 kcps) and 5.9 μW (2.0 μW) for the $E_x$ ($E_y$) transitions.

## 6- Impact of strain

The energy, cyclicity and strength of the NV optical transitions are impacted by the transverse electric or strain field in the crystal [5,10]. While low strain NVs such as NV B are preferable, most high strain NVs still display spin-selective transitions that can be isolated (see, e.g., Fig. 4 in Ref. [5]). NVs in the same crystal generally have a large distribution of transverse strain, which tends to be higher for NVs in nanostructures or near the surface. Figure S3(a,b) shows the PLE spectra of low strain NV B as well as higher strain NV C. We see that the relative strength of the $E_x$ and $E_y$ transitions is opposite and that the $m_S = 0$ cycling transitions ($E_x$ and $E_y$) are not as well separated from the $m_S = \pm 1$ transitions. The saturation curves for the $E_x$ and $E_y$ transitions (Fig. S3(e,f)) show that the maximum PL is much weaker than for the low strain NV B. This is, however, also the case for green excitation, with maximum CR of 13 kcps for a saturation power of 800 μW (cw illumination, curve not shown). As shown in Fig. S4, we perform the same optimization of the spin RO as for NV B but looking at both $E_x$ and $E_y$ transitions. The PL as a function of resonant illumination time for $E_x$ shows an intermediate behavior due to overlapping $E_x$ and $A_{12}$ transitions. Despite a lower spin contrast, we find this transition to be the most effective thanks to a higher saturation CR. We predict an SNR enhancement with red RO of 3.8 for $T_E = 2$ μs, which we verify by comparing green and red magnetic sensitivities under pulsed ODMR (measured ratio of sensitivities of 3.6). Note that additional measurements with higher-strain NV D show an enhancement of 4.02±0.60, which indicates a non-trivial dependence on strain.

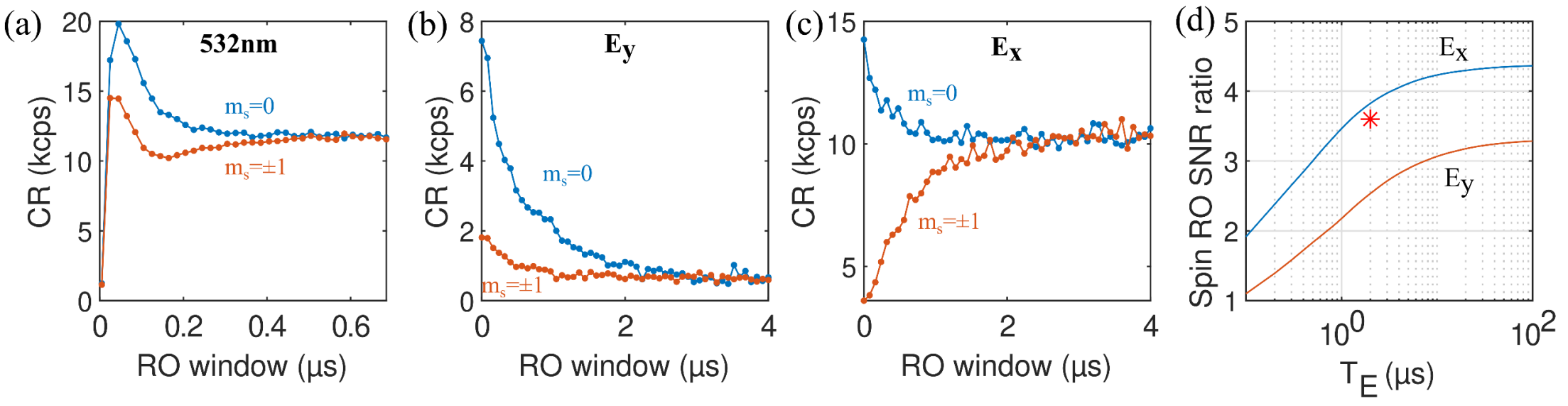

**Figure S4**: Evolution of the PL for NVC as a function of the illumination time under (a) 1.5 mW of green illumination and 30 µW of red illumination at the (b) $E_y$ or (c) $E_x$ wavelength. RO bins are 17 ns for (a) and 150 ns for (b,c). The spin is initialized with a green pulse and inverted with a 60 ns π-pulse. Time constants for the spin signal exponential decay are 120ns, 1µs and 0.6µs. (d) Expected SNR ratio between red and green RO for optimal RO window as a function of the spin evolution time. The red asterisk is an experimental measure of the magnetic sensitivity ratio for pulsed ODMR.

## 7-Measurement of the nuclear spin populations

Measurement of the nuclear spin population under red illumination is performed via a pulsed-ODMR of the NV electron spin under a frequency-selective π-pulse (typically > 2 µs) to resolve the $^{14}$N hyperfine structure. For each pulsed-ODMR, we measure a bright reference (PL under red illumination after charge initialization with the green laser) and use the fluorescence level from laser-detuned points to subtract the background photon count. The ODMR contrast is then calculated as the background-corrected PL signal in the presence of resonant MW minus the background-corrected reference. We then calculate the nuclear spin population in each projection as the ODMR contrast for the corresponding transition divided by the sum of the contrasts for all nuclear spin projections. Note that the sum of the contrasts typically adds up to below one and depends on the specific sequence used, which we attribute to ionization of the NV under red illumination or to residual populations in $|m_S = 0\rangle$ after spin-depletion. For example, in Fig. 3d of the main text the sum of the contrast from all six peaks gives a total of 0.77, 64% of which corresponds to the target state $|m_S = -1, m_I = -1\rangle$ and determines the protocol nuclear spin polarization efficiency.

## 8-Iterative nuclear spin hyperpolarization

Fig. 3c of the main text lays out the key hyperpolarization protocol, which relies on the non-cyclicity of the resonant optical transition to accumulate populations into a specific electron and nuclear spin state. Here, we use the $E_y$ transition, which is selective to the $|m_S = 0\rangle$ spin states but has a non-zero probability to relax into the $|m_S = \pm1\rangle$ states every optical cycle; as shown above, for every electron spin flip, there is an additional probability to flip the nuclear spin. In order to accumulate population in a specific electron-nuclear spin state (e.g., in $|m_S = -1, m_I = -1\rangle$), we iterate two-step sequences aimed at depleting all other populations: We first selectively address any undesired $|m_S = \pm1\rangle$ states with a MW π-pulse to transfer their populations into $|m_S = 0\rangle$ and then use the resonant laser to redistribute those populations into $|m_S = \pm1\rangle$. In the case presented in Fig. 3d of the main text, the MW pulses address all $|m_S = \pm1\rangle$ spin states except for $|m_S = -1, m_I = -1\rangle$, necessarily leading to accumulation of population in that state.

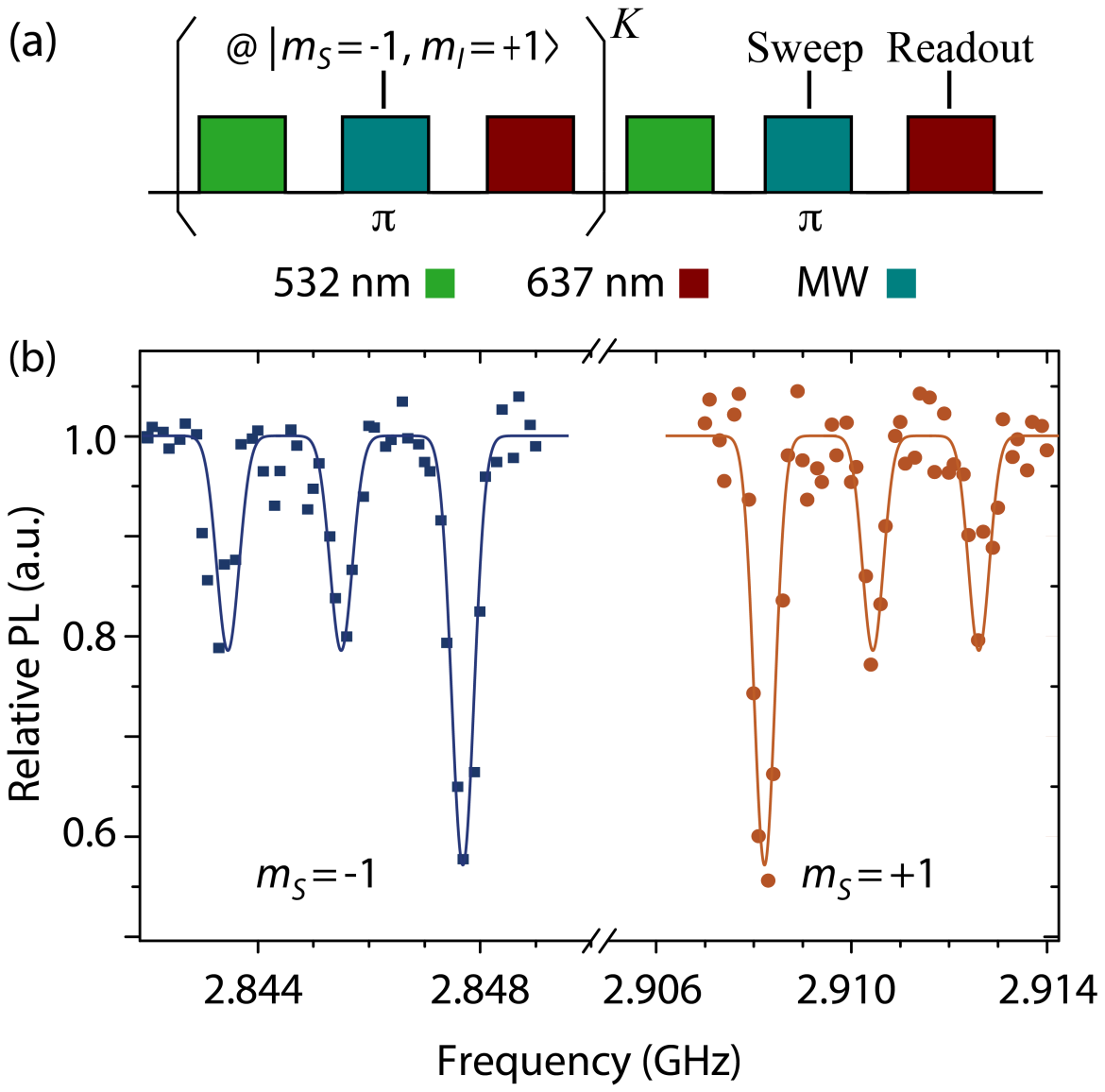

**Figure S5**: Probing the polarization occurring during a pulsed-ODMR sequence with weak spin re-initializations. (a) Experimental protocol; the first π-pulse is fixed to a specific transition, while the frequency of the second is swept across all transitions. (b) ODMR spectrum displaying hyperpolarization. Here $K = 5$ and the green pulse is 10 µW with a duration of 11 µs; solid lines are Lorentzian fits.

The maximum hyperpolarization we obtain of 68% can be explained by several factors. For one, we observe that for several NVs — including the one used for the hyperpolarization demonstration — a resonant laser tuned to the $E_y$ transition does not yield full depletion of the $|m_S = 0\rangle$ spin state, meaning that a non-zero PL is still observed after the initial spin-depletion decay. This could be due to proximity to the weaker $E_{1,2}$ transitions, and will lower the hyperpolarization in the limit of many depletion sequences since it will cause the hyperpolarized state to remain somewhat sensitive to resonant illumination. In practice, the overhead it creates also limits the number of iterations of depletion sequences we typically apply. Imperfect π-pulses can also render the spin pumping process slower and hence reduce the end nuclear spin polarization, a problem particularly relevant in the limit of only one or two depletion cycles.

| End state $\vert m_S, m_I\rangle$ | | Initial state $\vert m_S = 0, m_I = i_0\rangle$ | | |
|---|---|---|---|---|
| $m_I$ | $m_S$ | +1 | 0 | -1 |
| +1 | -1 | **0.28 (2)** | 0.06 (2) | 0.06 (2) |
| +1 | +1 | **0.38 (2)** | 0.06 (2) | 0.10 (2) |
| 0 | -1 | 0.09 (2) | **0.28 (2)** | 0.08 (1) |
| 0 | +1 | 0.06 (1) | **0.45 (2)** | 0.09 (2) |
| -1 | -1 | 0.14 (2) | 0.05 (2) | **0.28 (2)** |
| -1 | +1 | 0.05 (1) | 0.08 (2) | **0.39 (2)** |
| Total conserved fraction | | **0.66 (3)** | **0.66 (3)** | **0.74 (3)** |

**Table 2:** Populations distribution in the different electron-nuclear spin states after full electron spin depletion via resonant $E_y$ excitation, starting in $|m_S = 0, m_I = i_0\rangle$; values in the table represent the fraction of the population in the initial state that landed in the final state. Nuclear-spin-conserved fractions are highlighted in bold blue while double quantum transitions are shown in red.

**9-Measurement-induced nuclear spin polarization**

Another more time-efficient path to hyperpolarization is to use the measurement sequence itself to perform hyperpolarization. As exemplified in Fig. 4a of the main text, using a pulsed-ODMR sequence with only a partial spin-re-initialization allows accumulation of nuclear spin polarization as the sequence is repeated. Since unlike in the previous method we re-initialize the electron spin state with a green laser, all nuclear spin states that are not addressed by the microwave will be depleted and leak out into the nuclear spin state the microwave is resonant with. This mechanism relies on the green laser polarizing the electron spin faster than the nuclear spin and leads to a slightly reduced electron spin polarization. Figure 4a shows the pulsed ODMR response obtained with hyperpolarization happening for each of the different nuclear spin states at different frequencies, but we can also probe the hyperpolarization of a specific spin state by first iterating this sequence with a fixed microwave (here addressing $|m_S = -1, m_I = +1\rangle$) followed by pulsed ODMR with a varying MW frequency (see Fig. S5).

**10-Nuclear spin relaxation during resonant red excitation**

To characterize the interplay between resonant red illumination and the nuclear spin state, we use the protocol in Fig. 3a of the main text to measure the redistribution of population starting from the state $|m_S = 0, m_I = i_0\rangle$ (where $i_0 = 0, \pm 1$) into the six projections $m_S = \pm 1$, $m_I = 0, \pm 1$; the red laser pulse (20 μs, 1 μW) acts resonantly on the $E_y$ transition. To initialize the NV, we first trap the electron spin population in $|m_S = s_0\rangle$, $s_0 = \pm 1$ via a succession of red laser pulses and nuclear agnostic π-pulses acting in the $m_S = -s_0$ manifold, and subsequently apply a mw π-pulse selective on the $|m_I = i_0\rangle$ state, therefore preparing the NV in a statistical mixture of $|m_S = 0, m_I = i_0\rangle$ and $|m_S = s_0, m_I \neq i_0\rangle$. We finally deplete the $|m_S = 0, m_I = i_0\rangle$ state via a red $E_y$ laser pulse, and then probe the resulting population distribution by performing low power pulsed-ODMR on both $m_S = \pm 1$ multiplets. The top (bottom) plot in Fig. 3b shows the population after (before) the last two steps (nuclear-spin-selective π-pulse and red laser pulse) for $s_0 = -1, i_0 = -1$.

To probe the nuclear spin relaxation from any starting state $|m_I = i_0\rangle$, we used a simplified protocol taking five points of the pulsed ODMR, namely three resonant with the hyperfine structure and two off-resonance to serve as a reference for the background PL. For each initial nuclear spin projection $m_I = i_0$, we performed the measurement twice, initializing in $|m_S = s_0\rangle$ ($s_0 = \pm 1$) and reading out in the opposite electron spin state ($|m_S = -s_0\rangle$) so as to avoid reading background populations from the initial state $|m_S = s_0, m_I \neq i_0\rangle$. We also performed a readout before the last two steps to subtract any background populations in $|m_S = -s_0\rangle$ that could stem from imperfect initialization into $|m_S = s_0\rangle$.

Table 1 shows the resulting population for all three initial nuclear spin states. Interestingly, we find that the $|m_I = \pm 1\rangle$ states show efficient double quantum relaxation to the opposite nuclear spin state. Unlike the illustration in Fig. 3 of the main text, this statistical analysis reveals this double-quantum relaxation process is somewhat more efficient in the case where the end electron spin state shares the same sign; on the other hand, the $|m_I = 0\rangle$ state appears slightly more stable. We discuss the mechanisms governing the observed dynamics in the following section.

**11-Origin of the nuclear spin flip in the excited state**

In the $^3A_2$ ground state, both the electronic and $^{14}N$ nuclei spins are triplets, i.e., $I = S = 1$. The spin Hamiltonian may be expressed as

$$H = D\left(S_z^2 - \frac{1}{3}S(S+1)\right) + Q\left(I_z^2 - \frac{1}{3}I(I+1)\right) + A_\parallel S_z I_z + \frac{1}{2}A_\perp(S_+I_- + S_-I_+) + W. \quad (1)$$

The $^{14}N$ host sits in the symmetry axis of the defect, implying this spin Hamiltonian follows the $C_{3v}$ symmetry. In the ground state, $D^{(g)} = 2.87$ GHz, $Q^{(g)} = -4.96$ MHz, $A_\parallel^{(g)} = -2.16$ MHz, and $A_\perp^{(g)} \approx -2.70$ MHz [11]. Phonon relaxation processes are unlikely at cryogenic temperatures, so nuclear spin flips are rare and cannot be observed. Therefore, we focus on the spin Hamiltonian of the doubly degenerate $^3E$ excited state where $W$ takes the form

$$\begin{aligned} W = D_1[(S_zS_+ + S_+S_z)\sigma_- + (S_zS_- + S_-S_z)\sigma_+] + D_2[S_-^2\sigma_- + S_+^2\sigma_+] + \\ Q_1[(I_zI_+ + I_+I_z)\sigma_- + (I_zI_- + I_-I_z)\sigma_+] + Q_2[I_-^2\sigma_- + I_+^2\sigma_+] + \\ A_1[(I_zS_+ + I_+S_z)\sigma_- + (I_zS_- + I_-S_z)\sigma_+] + A_2[S_-I_-\sigma_- + S_+I_+\sigma_+]. \end{aligned} \quad (2)$$

In the above expression, $\sigma_\pm = |e_\mp\rangle\langle e_\pm|$ where $|e_\mp\rangle = (|e_x\rangle \pm i|e_y\rangle)/\sqrt{2}$ are the orbital flipping operators, and $D_j$, $Q_j$, and $A_j$, $j = 1,2$ denote coupling constants to the electron spin dipole-dipole, nuclear quadrupole, and hyperfine interactions. For future reference, we note that only the $Q_2$ term may double flip the $^{14}N$ nuclear spin.

Some parameters in Eq. (2) are known or deduced from experiment [10], e.g.,

$$D^{\text{expt}} = 1.42 \text{ GHz}, D_1^{\text{expt}} \propto 200/\sqrt{2} \text{ MHz}, D_2^{\text{expt}} \propto 1550/2 \text{ MHz},$$
$$A_\parallel^{\text{expt}} \approx 40 \text{ MHz}, A_\perp^{\text{expt}} \approx 27 \text{ MHz},$$

whereas others are not (e.g., $A_{1,2}^{\text{expt}}$). Note that a Jahn-Teller reduction factor ($q$) will appear in Eq. (2) via the dynamic Jahn-Teller effect [12,13]. Our *ab initio* results within PBE functional with $q \approx 0.5$ are

$$D = 1.67 \text{ GHz}, D_1 \propto -444 \times q \text{ MHz}, D_2 \propto 1382 \times q \text{ MHz},$$

in good agreement with experiment. On the other hand,the calculated ab initio quadrupole tensor is

$$Q \approx -3.9 \text{ MHz}, Q_1 \approx 27 \times q \text{ kHz}, Q_2 \propto 26 \times q \text{ kHz},$$

whereas the predicted hyperfine parameters are

$$A_\parallel \approx -40 \text{ MHz}, A_\perp \approx -23 \text{ MHz}, A_1 \approx 141 \times q \text{ kHz}, A_2 \approx 75 \times q \text{ kHz}$$

consistent with experiment.
The relaxation rate for the nuclear spin will then be

$$\Gamma_{0\to-1} = \frac{2\pi}{\hbar}\langle 0|\widehat{W}| - 1\rangle.$$

Initially, we excite the system from $|^3A_2\rangle \otimes |m_I\rangle$. Within the ground manifold, the hyperfine flipping is negligible as $A_\perp \approx \pm 2$ MHz and the quadrupole tensor does not have any flipping terms that can alter the $|m_I\rangle$ spin states. We excite the system with an optical photon to the $|^3E\rangle$ manifold. Those states are split into the $m_S = 0$ substates that are usually labeled by $|E_{x,y}\rangle = (|ae_{x,y}\rangle + |e_{x,y}a\rangle) \otimes (|\uparrow\downarrow\rangle - |\downarrow\uparrow\rangle)/2$ (for the sake of simplicity, the four $m_S = \pm 1$ substates labeled by $|A_{1,2}\rangle$ and $|E_{1,2}\rangle$ are not discussed in detail). Here we note that the $|E_{x,y}\rangle$ states split by a few GHz in the experiments. Now, we assume that before and during the early stages of the optical excitation the nuclear spin cannot flip, thus leading to initialization into $|E_x\rangle \otimes |m_I\rangle$. However, this state is perturbed, for example, by $W = \cdots + Q_1[(I_zI_+ + I_+I_z)\sigma_- + \cdots$. First, we evaluate the effect of the $\sigma_\pm$ orbital operators,

$$\begin{pmatrix} \langle E_x| \\ \langle E_y| \end{pmatrix} \sigma_\pm (|E_x\rangle |E_y\rangle) = \frac{1}{2}\begin{pmatrix} 1 & \pm i \\ \mp i & 1 \end{pmatrix} = \frac{1 \pm \sigma_y}{2} \quad (3)$$

Next, we need to evolve the system; Fermi's golden rule only applies if $\frac{4|W_{fi}|^2}{(E_f - E_i)^2} \ll 1$, e.g., it cannot be applied to $|E_x\rangle \otimes |m_I\rangle \leftrightarrow |E_x\rangle \otimes |m_I'\rangle$ degenerate states.

We first initialize into the $|E_x\rangle \otimes |0\rangle = |e_x^0\rangle \otimes |0\rangle$ substate where the $^0$ superscript denotes the $m_s = 0$ substate. The $V = \frac{1}{2} A_\perp S_+ I_-$ hyperfine gives the following perturbation.

$$\widetilde{|E_x\rangle \otimes |0\rangle} = |E_x\rangle \otimes |0\rangle + \frac{A_\perp}{E(E_x) - E(E_1)} (|E_x\rangle \otimes |x\rangle - |E_2\rangle \otimes |y\rangle), \quad (4)$$

where $|\pm\rangle = (|x\rangle \pm i|y\rangle)/\sqrt{2}$ and $E(E_x) - E(E_1) \approx 7$ GHz (due to strain, see Fig. 1. in main text). Therefore, the level anti-crossing (LAC) rate will be as follows: $\Gamma_{A_\perp} = A_\perp^2 / \sqrt{A_\perp^2 + (7\text{ GHz})^2} \approx \frac{A_\perp^2}{7\,GHz} \approx 100$ kHz. First, it seems that this will be the primary source of nuclear spin flipping, since all other $Q_{1,2}$, $A_{1,2}$ flipping terms are an order of magnitude smaller than that for $A_\perp$. However, there is a caveat: If we induce the $|E_x\rangle \otimes |m_I\rangle \leftrightarrow |E_x\rangle \otimes |m_I'\rangle$ transition, then the energy separation is significantly smaller as only the quadrupolar interaction splits these states. Therefore, we directly induce a transition with $Q_{1,2}$ operators that avoids electron spin flipping, for instance, via $V = qE_1 I_+ I_z \sigma_-$, i.e.,

$$\widetilde{|E_x\rangle \otimes |0\rangle} = |E_x\rangle \otimes |0\rangle + \frac{\sqrt{2} q Q_1}{Q} |E_x\rangle \otimes |x\rangle. \quad (5)$$

Therefore, the LAC rate will be $\Gamma_{Q_1} = \frac{2q^2 Q_1^2}{3.9\text{ MHz}} \approx 46$ Hz and can be neglected. However, if we do the same for the $m_I = \pm 1$ states not separated by $Q$ splitting by means of $V = Q_2 I_\mp^2 \sigma$, then

$$V|E_+\rangle \otimes |-\rangle = 2qQ_2 |E_-\rangle \otimes |+\rangle, \text{ or equivalently } V|E_x\rangle \otimes |x\rangle = -2qQ_2 |E_y\rangle \otimes |y\rangle \quad (6)$$

where $|E_\pm\rangle = (|E_x\rangle \pm i|E_y\rangle)/\sqrt{2}$. Therefore, the LAC rate will be $\Gamma_{Q_2} = 2qQ_2 \approx 26$ kHz. Note that if an external magnetic field splits $|\pm\rangle$, the rate will be slower as follows

$$\Gamma_{Q_2} = \frac{(2qQ_2)^2}{\sqrt{(2\gamma_I B_z)^2 + (2qQ_2)^2}} \approx 24\text{ kHz} \quad (7)$$

where the magnetic field is set to $B_z = 1.2$ mT, and $\gamma_I = 431.7$ Hz/Gauss for $^{14}$N, thus $2\gamma_I B_z = 10.3$ kHz.

In summary, we find that

- $|m_I\rangle \leftrightarrow |m_I'\rangle$ flipping is done by means of the transverse $A_\perp$ hyperfine coupling. This interaction flips the electronic spin too, so the energy differences between the $|A_{1,2}\rangle$, $|E_{1,2,x,y}\rangle$ substates do matter. Therefore, the hyperfine interaction mediated rate $\Gamma_{hyper} \approx \frac{A_\perp^2}{7\,GHz} \approx 100$ kHz. This process flips $I$ only by one quantum, i.e., $\Delta m_I = \pm 1$. However, it is independent on which $|A_{1,2}\rangle$,$|E_{1,2,x,y}\rangle$ we optically pump the system into, it is always present because $A_\perp = 27$ MHz is very effective in inducing $m_I$ flipping interactions.
- However, for $m_S = 0$ of $|E_{x,y}\rangle$ substates new routes open up, i.e., one may double flip $I$ ($\Delta m_I = \pm 2$) via $Q_2[I_-^2 \sigma_- + I_+^2 \sigma_+]$. Since the value of $Q_2$ is comparable to that of the Zeeman splitting of $^{14}$N nuclear spin, a resonance occurs that leads to a level anti-crossing (LAC) for the nuclear spins. Finally, the $\Gamma_{Q_2} \approx 24$ kHz transition occurs for $B_z = 1.2$ mT but only within the $|E_{x,y}\rangle$ manifold. We note that this process does not occur for any $m_S = \pm 1$ substates.

---